\documentclass[showpacs,preprintnumbers,amsmath,amssymb,twocolumn,nofootinbib]{revtex4} %nofootinbib

\usepackage{enumitem} 
\usepackage{dcolumn}% Align table columns on decimal point
\usepackage{bm}% bold math
\usepackage[dvips]{graphicx}
\usepackage{amsmath}
\usepackage{epsfig}
\usepackage{amsfonts}
\usepackage{amssymb}
\newcommand{\be}{\begin{equation}}
\newcommand{\ee}{\end{equation}}
\newcommand{\bee}{\begin{equation*}}
\newcommand{\eee}{\end{equation*}}
\newcommand{\bea}{\begin{eqnarray}}
\newcommand{\eea}{\end{eqnarray}}
\newcommand{\bean}{\begin{eqnarray*}}
\newcommand{\eean}{\end{eqnarray*}}

\newcommand{\nn}{\nonumber}

\usepackage{color}

\begin{document}
\preprint{ULB-TH/12-23 USM-TH-310}

%\mbox{ } \\\vspace{-3mm}
\title{\Large\mbox{ } \\\vspace{0mm}Predictive Model for Radiatively Induced Neutrino Masses \\and Mixings with Dark Matter}
%\title{\Large\mbox{ } \\\vspace{5mm}The Cocktail Model: Neutrino Masses and Mixings with Dark Matter\\\vspace{1mm}}
%\title{\Large\mbox{ } \\\vspace{5mm}Cocktail Model for Neutrino Masses and Mixings with Dark Matter\\\vspace{1mm}}

\author{Michael Gustafsson,$^{1}$ Jose M. No,$^{2}$ and Maximiliano A. Rivera$^{3}$\vspace{1mm}}
\affiliation{$^{1}$ Service de Physique Th\'eorique, Universit\'e Libre de Bruxelles, B-1050 Bruxelles, Belgium}
\affiliation{$^{2}$ Department of Physics and Astronomy, University of Sussex, BN1 9QH Brighton, United Kingdom}
\affiliation{$^{3}$ Departamento de F\'isica, Universidad T\'ecnica Federico Santa Mar\'ia, Casilla 110-V, Valparaiso, Chile\vspace{0mm}}

%\author{Michael Gustafsson}
%\email{mgustafs@ulb.ac.be}
%\affiliation{Service de Physique Th\'eorique, Universit\'e Libre de Bruxelles, B-1050 Bruxelles, Belgium}
%
%\author{Jose M. No}
%\email{J.M.No@sussex.ac.uk}
%\affiliation{Department of Physics and Astronomy, University of Sussex, BN1 9QH Brighton, United Kingdom}
%
%\author{Maximiliano A. Rivera}
%\email{maximiliano.rivera@usm.cl}
%\affiliation{Departamento de F\'isica, Universidad T\'ecnica Federico Santa Mar\'ia, Casilla 110-V, Valparaiso, Chile}

\date{May 16, 2013\\\vspace{3mm}}% It is always \today, today,

\begin{abstract}
A minimal extension of the standard model to naturally generate small neutrino masses and provide a dark matter  candidate is proposed. The dark matter particle is part of a new scalar doublet field that plays a crucial role  in radiatively generating neutrino masses. The symmetry that stabilizes the dark matter also suppresses  neutrino masses to appear first at three-loop level. Without the need of right-handed neutrinos or other very heavy  new fields, this offers an attractive explanation of  the hierarchy between the electroweak and neutrino mass scales. The model has distinct verifiable  predictions for the neutrino masses, flavor mixing angles, colliders and dark matter signals.
\end{abstract}

\pacs{14.60.Pq, 12.60.Fr, 95.35.+d \vspace{1mm}}
\maketitle

%%%%%%%%%%%%%%%%%%%%%%%%%%%%%%%%%%%%%%%%%%%%%%%%%%%%%%%%%%%%
%\section{Introduction.}
%\vspace{-3mm}
%%%%%%%%%%%%%%%%%%%%%%%%%%%%%%%%%%%%%%%%%%%%%%%%%%%%%%%%%%%%
The existence of a large amount of nonbaryonic dark matter in the Universe and the  observation of  nonzero neutrino masses may be regarded as the most direct and compelling evidence of particle  physics beyond the standard model (SM). However, both the origin of neutrino masses and  the nature of dark matter  are still unknown. 
%\footnote{\red{For the allowed range of neutrino masses, the neutrinos can  themselves only contribute a negligible amount of the observed DM abundance.}}
A scenario of being able to  incorporate both phenomena in a  unified framework would then be very attractive.

One of the best motivated dark matter scenarios is that of stable weakly interacting massive  particles (WIMPs), produced as a thermal relic from the early Universe.  Among the simplest realizations of  this WIMP scenario is the inert doublet model \cite{Ma:2006km,IDM}. The SM is extended by a scalar doublet $\Phi_2$, and the dark matter scalar is made stable due to an exact $Z_2$ symmetry under which the new field has odd parity.  The inert doublet model is currently constrained by results from dark matter searches as well as by particle collider data, but still a large region of its parameter space is allowed.

From the perspective of neutrino physics, currently the most popular way to generate small neutrino masses  is the see-saw mechanism (see \cite{seesaw} for a review). In its simplest variant, it postulates the existence of very massive $SU(2)_{L}\times U(1)_{Y}$ singlet right-handed  neutrinos. Other realizations involve the existence of very  heavy scalar or  fermionic triplets. Although elegant, this mechanism is difficult (if not impossible) to test, as the masses of the new states are typically much larger than can be experimentally probed. 

Small neutrino masses can also be generated via radiative corrections. This has been explored  by explicit lepton number violation in extensions of the scalar sector of the SM. As opposed to  the see-saw mechanism, this approach generates small neutrino masses without relying on  new particles at a very high energy scale.  One simple realization of this idea is the Zee model \cite{Zee:1980ai}, in which the SM field content  is enlarged by a second scalar doublet $\Phi_2$ and a charged scalar singlet $S^{+}$. Another simple scenario  is the Zee-Babu model \cite{Babu:1988ki}, which replaces the scalar doublet $\Phi_2$ in the Zee model by a  doubly charged singlet scalar field $\rho^{++}$. 

However, these scenarios of radiative neutrino mass generation in \cite{Zee:1980ai}  and \cite{Babu:1988ki} (together with many others, such as \cite{Nasri:2001ax,FAguila}) cannot at  the same time contain a viable dark matter candidate. A stabilizing symmetry for some of the new fields would in fact forbid the very terms responsible for the generation of neutrino  masses (see, however, \cite{Ma:2006km,Krauss:2002px} for interesting scenarios including  right-handed neutrinos and dark matter particle candidates).  This is for example the case in the Zee model if an odd $Z_2$ parity is assigned to $\Phi_2$.  Finding a unified scenario for radiative neutrino mass generation and a dark matter particle candidate is then a nontrivial task.

In this Letter we construct a minimal model, that generates neutrino masses radiatively and provides a stable dark matter candidate, via an extended scalar sector with an exact $Z_2$ symmetry. We do this by adding to the SM two scalar singlets, $\rho^{++}$ and $S^{+}$, and a scalar doublet 
$\Phi_2$ with masses around the electroweak (EW) scale. The fields $S^{+}$ and $\Phi_2$ have odd $Z_2$ parity (while all other fields do not transform under this symmetry), and therefore a variation of the mentioned inert doublet model of dark matter is automatically embedded into the scenario. Due to the $Z_2$ symmetry and the field content of the model, Majorana neutrino masses are first generated at the three-loop level, naturally explaining the large hierarchy $m_{\nu}/v \sim 10^{-13}$ as due to the loop suppression  $(g^2/16\pi^2)^{3} \sim 10^{-13}$  ($g$ being an EW-sized coupling and with all masses at the EW scale $v$) \cite{Farzan:2012ev}. This scenario  then provides an intrinsic and interesting link between the stability of the dark matter candidate and the smallness of the neutrino mass scale.

%\medskip
%The Letter is organized as follows: in section~\ref{sec:I} we introduce the model, obtain the neutrino mass 
%matrix and verify that it can produce the observed pattern of neutrino masses and mixings. In section~\ref{sec:II} we 
%discuss the properties of the dark matter candidate. In section~\ref{sec:III} we analyze the constraints on  the model coming 
%from lepton flavor violating processes, EW precision tests (EWPTs) and collider searches, and  comment on possible 
%consequences for neutrinoless double beta decay.

%%%%%%%%%%%%%%%%%%%%%%%%%%%%%%%%%%%%%%%%%%%%%%%%%%%%%%%%%%%%
\section{A Model for Neutrino Masses.}\label{sec:I}
%\vspace{-2mm}
%%%%%%%%%%%%%%%%%%%%%%%%%%%%%%%%%%%%%%%%%%%%%%%%%%%%%%%%%%%%
In addition to the SM fields, the model includes two $SU(2)_{L}$ singlet  scalars (singly and doubly charged) $S^{+}$ and $\rho^{++}$, and a scalar doublet $\Phi_2$.  We introduce a $Z_2$ symmetry under which the $\Phi_2$ and $S^{+}$ fields are odd,  whereas $\rho^{++}$ and the SM fields are even.  The $Z_2$ symmetry should be unbroken after EW symmetry breaking, so that the lightest  $Z_2$-odd state remains stable and can provide a dark matter particle candidate. Given the symmetry and particle content of the model, the Lagrangian will include the following  relevant terms leading to lepton number violation: 
\bea
\label{LGNR}
-\Delta \mathcal{L} = \frac{\lambda_5}{2} \left(\Phi^{\dagger}_{1} \Phi_2 \right)^2 
+ \kappa_1 \, \Phi_2^{T} i \sigma_2 \Phi_{1}\, S^{-}  + 
\kappa_2 \, \rho^{++} S^{-} S^{-} \nonumber \\
+ \xi \, \Phi_2^{T} i \sigma_2 \Phi_{1}\, S^{+} \, \rho^{--} 
 + C_{ab} \, 
\overline{\ell^{c}_{aR}} \ell_{bR} \, \rho^{++} +  \mathrm{H.c.} \quad \;\;
\eea
Here, $a,b$ denote family indices of the right-handed charged leptons $\ell_R$, and  the Yukawa couplings $C_{ab}$ form a symmetric and complex matrix, allowing for charge-parity ($CP$) violation in the leptonic sector.

The SM scalar doublet $\Phi_1$ and the inert scalar doublet $\Phi_2$ can in the unitary gauge be written as
\be
\Phi_1 = \frac{1}{\sqrt{2}}\left(\begin{array}{c}
0 \\
h
\end{array}
\right) 
+
\left(\begin{array}{c}
0 \\
v
\end{array}
 \right),
\, \, \Phi_2 =\frac{1}{\sqrt{2}} \left(
\begin{array}{c}
\Lambda^{+} \\
H_0 + i\, A_0
\end{array} \right),
\ee
where $v \simeq 174$ GeV is the vacuum expectation value of $\Phi_1$.  After  EW symmetry breaking, and for $\kappa_1 \neq 0$, the charged states
$\Lambda^+$ and $S^+$ will mix (the mixing angle being $\beta$), giving rise to two charged mass eigenstates 
\be
 H^+_1 = s_{\beta}\, S^+ + c_{\beta}\, \Lambda^+ , \quad \quad 
 H^+_2 =  c_{\beta}\, S^+ -  s_{\beta}\, \Lambda^+,
\ee
with $s_\beta, c_\beta = \sin\beta, \cos\beta$, respectively.
A convenient set of independent variables may be given by  the five new scalar masses $m_{\rho,H^0,A^0,H_{1,2}^\pm}$, the mixing  angle $\beta$ and the couplings $\xi$ and $\kappa_{2}$. All coefficients in the  scalar potential should be chosen within their perturbative regime and to make the potential preserve vacuum stability \cite{Babu:2002uu} and fulfil dark matter  constraints.

\begin{figure}[t]
\center{\includegraphics[width=0.92 \columnwidth]{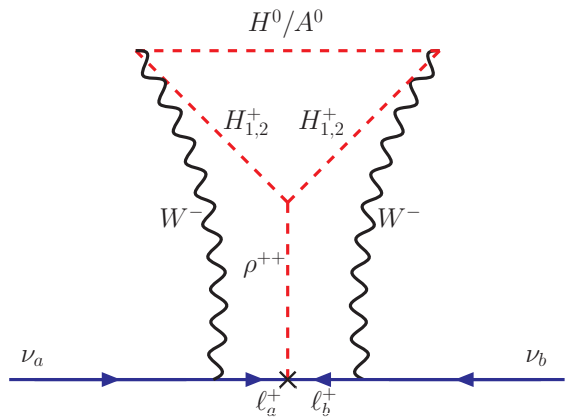}}
\caption{
The ``cocktail diagram.''}
%The `Cocktail diagram', giving rise to neutrino mass generation.}
%The Feynman `Cocktail diagram'.} %responsible for the neutrino mass generation.} 
\label{fig:Cocktail}
\end{figure}

\medskip
The Lagrangian in Eq.\;(\ref{LGNR}) breaks lepton number explicitly  by two units \footnote{The lepton number is conserved if either $C_{ab}$ or $\lambda_5$ vanish. If $\kappa_1 = 0$, the lepton number is still violated for  nonvanishing $\kappa_2$ and $\lambda_{\rho S}$, but the leading contribution  to neutrino masses appears at the five-loop level.},  which generates a Majorana mass for the left-handed neutrinos. With the viable matter content neutrino masses can never be generated at one-loop order and the $Z_2$ symmetry precisely  forbids all terms that would have generated neutrino masses at  two-loop order. Therefore the leading contributions to neutrino masses appear first at three loops -- through the ``\textit{cocktail diagram}'' shown in Figure~\ref{fig:Cocktail}. 
%
%
%%%%%%%%%%%%%%%% 
%%%%%%%%%%%%%%%%
%
%

In the basis where charged current interactions are flavor diagonal and the charged leptons $e, \mu, \tau$ are mass eigenstates, the summed contributions of the six different finite three-loop diagrams shown in Figure~\ref{fig:Cocktail}  (coming from $H^{+}_{1,2}$, $A_0$ and $H_0$ running in the loop) give the Majorana neutrino mass matrix: 
\be
\label{MajoranaMatrix}
m^{\nu}_{ab} = C_{ab} x_a x_b  \frac{s_{2\beta}}{(16\pi^2)^3} \, (\mathcal{A}_1 \, \mathcal{I}_1 + \mathcal{A}_2 \, \mathcal{I}_2), 
\ee 
where $s_{2\beta} = \sin(2\beta)$, $x_a = m_a/v$ for $a = e, \mu, \tau$, and
\bea
\mathcal{A}_1 &\simeq& 
 \frac{5 M_W^4}{v^6} (\Delta m_{+}^2)^2 \Delta m_{0}^2
\times
\frac
{ \kappa_2 s_{2\beta} + \xi v \, c_{2\beta} }
{ \;\,  m_\rho^{3/2}\, m_0^{1/2}\, m_+^{2\phantom{/1}}}, \nonumber
\\
\mathcal{A}_2 &\simeq& 
-\frac{29 M_W^4}{v^6} 
\Delta m_{+}^2 \Delta m_{0}^2 
\times 
\frac{\xi v}
{ \; m_\rho\, m_0^{1/2}\, m_+^{1/2}}.\nonumber
\eea
The notations introduced are that $\Delta m^2_{0}\!=\!m^2_{A_0}\!-m^2_{H_0}$ and $\Delta m^2_{+}\!=\!m^2_{H_2^+}\!-m^2_{H_1^+}$, 
as well that $m^2_{0}\!=\!m^2_{A_0}\!+m^2_{H_0}$ and $m^2_{+}\!=\!m^2_{H_2^+}\!+m^2_{H_1^+}$. 
The factors $\mathcal I_{1,2}$ are  dimensionless $\mathcal{O}(1)$ numbers emerging from  three-loop integrals after generic factors have been factored out. Their exact values depend on the specific mass  spectrum, and we have made the full 3-loop calculation \cite{Erratum} with the help of {\sf SecDec} \cite{SecDec} for numerical integrations. 
The normalization and the specific mass scaling with $m_\rho$, $m_0$ and $m_+$ in the above equations are found empirically; determined by fits to scans over model parameters around our benchmark point (presented below). This neutrino mass formula is typically accurate to within 30\;\% when input parameters are allowed to vary up to at least a factor of a few from the benchmark point.

The proportionality  of $m^{\nu}_{ab}$ to the mass differences $\Delta m^2_{0}$ and  $\Delta m^2_{+}$ signals a {Glashow-Iliopoulos-Maiani-like  (GIM-like) mechanism \cite{Glashow:1970gm} at play in Eq.\;(\ref{MajoranaMatrix}), which can be easily understood noticing that $\Delta m^2_{0} \propto \lambda_5$ and  $\Delta m^2_{+} \propto \kappa_1$. In the limit $\lambda_5 \rightarrow 0$  the Lagrangian in Eq.\;(\ref{LGNR}) conserves the lepton number and no Majorana neutrino mass can be generated, while in the limit $\kappa_1 \rightarrow 0$, the leading contribution to $m^{\nu}_{ab}$ will appear at a higher loop order. 

\medskip
We now analyze the ability of the model to reproduce the observed pattern of neutrino masses and mixings. The standard parametrization for the neutrino mass matrix in terms of three masses $m_{1,2,3}$,  three mixing angles $\theta_{12}$, $\theta_{23}$, $\theta_{13}$, and three phases $\delta$, $\alpha_1$, $\alpha_2$ reads
\be
\label{UPMNS}
m^{\nu} = U^{T} \,m^{\nu}_{D} \,U \quad \mathrm{with} \quad m^{\nu}_{D} = \mathrm{Diag}\left(m_1,m_2,m_3\right),
\ee
\vspace{-6mm} 
\bea
\label{UPMNS2}
U = \mathrm{Diag}\left(e^{i\alpha_{1}/2},e^{i\alpha_{2}/2},1\right) \times \quad \quad \quad 
\quad \quad \quad \nonumber\\ 
\left(
%\!\!\begin{array}{ccc}
\begin{matrix}
c_{13}c_{12} & -c_{23}s_{12}\!-\!s_{23}c_{12}s_{13}e^{i\delta} &   s_{23}s_{12}\!-\! c_{23}c_{12}s_{13}e^{i\delta} \\    
c_{13}s_{12} &  c_{23}c_{12}\!-\!s_{23}s_{12}s_{13}e^{i\delta} &  -s_{23}c_{12}\!-\!c_{23}s_{12}s_{13}e^{i\delta} \\
s_{13}e^{-i\delta} & s_{23}c_{13}  & c_{23}c_{13}
 \end{matrix}
%\end{array} \!\!
\right) 
 \nonumber
\eea
where $s_{ij} \equiv \sin(\theta_{ij})$ and $c_{ij} \equiv \cos(\theta_{ij})$.  A global fit to neutrino oscillation data after the recent measurement of $\theta_{13}$  (see for example \cite{NeutrinoData}) gives  $\Delta m^2_{21} \equiv m^2_{2} - m^2_{1} = 7.62^{+0.19}_{-0.19}\times 10^{-5} \mathrm{eV}^2$, $\left|\Delta m^2_{31}\right| \equiv \left| m^2_{3} - m^2_{1} \right|= 2.55^{+0.06}_{-0.09}\times 10^{-3}  \mathrm{eV}^2$, $s_{12}^2 = 0.320^{+0.016}_{-0.017}$, $s_{13}^2 = 0.025^{+0.003}_{-0.003}$, and $s_{23}^2 = 0.43^{+0.03}_{-0.03}$ ($0.61^{+0.02}_{-0.04}$) if in the first (second) octant for $\theta_{23}$. Neutrino oscillations are not sensitive to the Majorana phases $\alpha_1$ and $\alpha_2$ nor  to the absolute neutrino  mass scale, while the value of the $CP$ phase $\delta$ is beyond current  experimental sensitivity.  In the inverted hierarchy scenario ($\Delta m^2_{31} < 0$) these experimental data lead to $\left|m^{\nu}_{ee}\right| \gtrsim 10^{-2}$ eV, which cannot  be accommodated by Eq.\;(\
ref{MajoranaMatrix}) due to the $x_e^2 \sim 10^{-9}$ suppression of an already three-loop suppressed EW-sized mass scale $\mathcal{A}$. Thus, a normal hierarchy pattern for neutrino masses is predicted.

The fact that the entries $m^{\nu}_{ee,e\mu}$ in Eq.\;(\ref{MajoranaMatrix})  are parametrically much smaller than the rest (being proportional to $x_e^2$ and $x_e x_{\mu}$) results in an approximate neutrino mass texture of the form $m^{\nu}_{ee} \sim 0$, $m^{\nu}_{e\mu} \sim 0$.  For given values of $\Delta m^2_{21}$, $\Delta m^2_{31}$, $s_{12}^2$ and $s_{23}^2$, the four constraints $\mathrm{Re}\left[m^{\nu}_{ee} \right] , \mathrm{Im}\left[m^{\nu}_{ee} \right] \simeq 0$ and $\mathrm{Re}\left[m^{\nu}_{e\mu} \right] , \mathrm{Im}\left[m^{\nu}_{e\mu} \right] \simeq 0$ can in fact only be satisfied over a certain range of $s_{13}^2$ \cite{FAguila}. If $\Delta m^2_{21}$, $\Delta m^2_{31}$, $s_{12}^2$ and $s_{23}^2$ are taken at their central values of the global fit in \cite{NeutrinoData}, then the model predicts $0.009 \leq s_{13}^2 \leq 0.015$  ($s_{13}^2 \geq 0.017$) for $\theta_{23} > \pi/4$ ($\theta_{23} < \pi/4$). It thus gives a correlated prediction for $\theta_{13}$ and the deviation of $\theta_{23}$ 
from $\pi/4$ (maximal mixing) towards lower or larger values.  Moreover, for fixed values of $\Delta m^2_{21}$, $\Delta m^2_{31}$, $s_{12}^2$, $s_{23}^2$  and $s_{13}^2$ (and when allowed by the mass texture) the above constraints  lead to a specific prediction for $m_1$, $\alpha_{1}$,  $\alpha_{2}$ and $\delta$.

For the mass texture $\left|m^{\nu}_{ee,e\mu}\right| \simeq 0$ discussed above, the experimental pattern of neutrino masses and mixings  results in a neutrino mass matrix with the accompanying structure
$\left|m^{\nu}_{e\tau}\right| \simeq 1\times10^{-2}$\;eV, 
$\left|m^{\nu}_{\mu\mu,\mu\tau,\tau\tau}\right| \simeq 2.5\times10^{-2}$\;eV. 
From Eq.\;(\ref{MajoranaMatrix}), we find that in order to radiatively generate $m^{\nu}_{e\tau}$ and  $m^{\nu}_{\mu\mu}$ entries of the right size (bounds on  $C^{\nu}_{\mu\tau}$ and $C^{\nu}_{\tau\tau}$ are weaker) we need model parameters such that 
\bea
\label{Constrainnumass}
C_{e\tau} \, s_{2\beta}\, ({\cal A}_1 + {\cal A}_2) \simeq 1.3 \, \mathrm{TeV}, 
\\ 
C_{\mu\mu} \, s_{2\beta}\, ({\cal A}_1 + {\cal A}_2) \simeq 0.3 \, \mathrm{TeV}.
\eea

%%%%%%%%%%%%%%%%%%%%%%%%%%%%%%%%%%%%%%%%%%%%%%%%%%%%%%%%%%%%
\section{Dark Matter.}\label{sec:II}
%\vspace{-2mm}
%%%%%%%%%%%%%%%%%%%%%%%%%%%%%%%%%%%%%%%%%%%%%%%%%%%%%%%%%%%%
When the lightest $Z_2$-odd states is electrically neutral the model has a WIMP dark matter candidate. For the remaining of the paper this particle will be assumed to be $H_0$ (taking $A_0$ would be equivalent).  This WIMP scenario resembles the inert doublet model \cite{IDM} and should share much of its phenomenology  (see e.g.\ \cite{Gustafsson:2010zz} and references therein). 

The relic abundance of $H_0$ is determined by its annihilation rate at freeze-out. In the mass  range $m_{H_0}=50 - 75$~GeV \cite{Gustafsson2012} or above 520 GeV \cite{Hambye:2009pw}, the correct dark  matter abundance  can be achieved while being compatible with existing bounds from the Large Electron-Positron collider (LEP), electroweak precision tests (\mbox{EWPTs}),  and direct and indirect dark matter searches. The lower WIMP mass range allows us to simultaneously 
generate neutrino masses of the right size in our model.  The correct dark matter abundance can be reached in the following situations:
\begin{enumerate}[topsep=3pt, partopsep=1pt,itemsep=1pt,parsep=1pt,leftmargin=1.5em]
\item[(a)]  Annihilations into fermions  via resonant SM scalars when $m_{H_0} \sim m_h/2$. 
\item[(b)] Coannihilation with either $A_0$ or $H^{+}_{1,2}$, if the mass splitting  to $H_0$ is less than some GeV. For fairly large mass splittings  $\Delta m^2_{+}$ and  $\Delta m^2_{0}$, coannihilations $H_0$-$A_0$ are strongly suppressed, while coannihilations $H_0$-$H^+_{1}$ may still be possible. 
\item[(c)]  The WIMP mass approaching $m_W$, where the closeness to the $WW$ threshold regulates the annihilation rate at freeze-out.
\end{enumerate}

Apart from a potential signal in {direct dark-matter search experiments, the model could produce a striking monochromatic gamma-ray line \cite{Gustafsson:2007pc} detectable by the Fermi Large Area Telescope.

%%%%%%%%%%%%%%%%%%%%%%%%%%%%%%%%%%%%%%%%%%%%%%%%%%%%%%%%%%%%
\section{Experimental Constraints.}\label{sec:III}
%\vspace{-2mm}
%%%%%%%%%%%%%%%%%%%%%%%%%%%%%%%%%%%%%%%%%%%%%%%%%%%%%%%%%%%%
Direct searches  at LEP for doubly charged scalars $\rho^{++}$ decaying into same-sign dileptons set a  lower bound $m_{\rho} \gtrsim 160$ GeV \cite{LEP1} (which however depends on the value of $C_{ee}$). Bounds from virtual $\rho^{++}$ exchange in Bhabha scattering lead to $C^2_{ee} \lesssim 9.7 \times 10^{-6}\, \mathrm{GeV}^{-2} \, m_{\rho}^2$ \cite{LEP1,LEP2}. More stringent limits from direct searches at the Tevatron and the Large Hadron Collider (LHC) are more subtle to derive (as opposed to doubly charged scalars $\Delta^{++}$ from inside $SU(2)_L$ triplets, $\rho^{++}$ does not couple to $W$ bosons).  The ATLAS collaboration at the LHC searched for pair produced $\rho^{++}$ decaying into leptons and set a limit $m_{\rho} \gtrsim 400$ GeV \cite{ATLAS:2012hi}. For the charged states $H_{1,2}^+$, the LEP data from chargino searches can be translated into an approximate bound $m_{H^+} \gtrsim 70 - 90$ GeV  (depending on $m_{H_0}$) \cite{Pierce:2007ut}. Moreover,  LEP excludes models with $m_{A_0} \lesssim 
100$~GeV if $m_{H_0}\lesssim80$~GeV and $m_{A_0}-m_{H_0} \gtrsim 10$~GeV \cite{Lundstrom:2008ai}.

The new inert fields also contribute to EWPT observables, such as the oblique parameters $S$, $T$ and $U$  \cite{Grimus:2008nb}. For $m_h \simeq 126$ GeV, the most important constraint  is given by $\Delta T \in \left[ -0.04, 0.12 \right]$ at $95\%$ C.L. \cite{Baak:2012kk} (contributions to $S$ and $U$ are found to be negligible). The one-loop contribution to $T$ from the new fields is calculated to be 
\bea
\label{deltaT}
\Delta T = \frac{1}{16\pi m_W^2 s_{\theta_W}^2} 
\Big[ 
c_\beta^2 \left(F_{H_1^+,H_0} +F_{H_1^+,A_0}\right) \nn\\
+ s_\beta^2 \left( F_{H_2^+,H_0} +F_{H_2^+,A_0} \right) 
- 2 c_\beta^2 s_\beta^2 F_{H_1^+,H_2^+}-F_{H_0,A_0}\Big] 
\eea
with $F_{i,j} = \frac{m_i^2+m_j^2}{2}-\frac{m_j^2 m_j^2}{m_i^2-m_j^2} \ln{\frac{m_i^2}{m_j^2}}$, and $\theta_W$ being the Weinberg angle. EWPT constraints can be 
satisfied for a wide range of masses and mixing angles $\beta$ and, as opposed to the inert doublet model \cite{IDM},  the present scenario allows for large mass splittings (see Figure \ref{fig2:Cocktail}). 
%
%%%%%%%%%%%%%%%% 
\begin{figure}[t!]
\center{\includegraphics[width=0.9 \columnwidth]{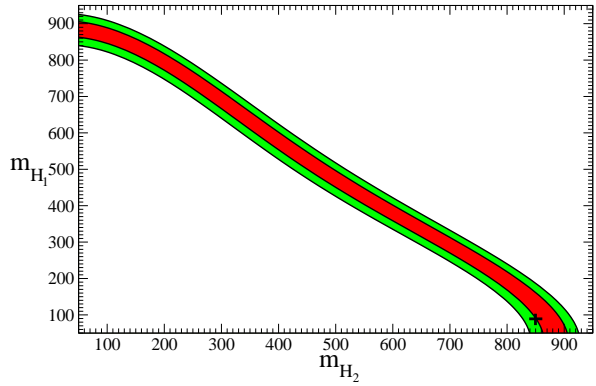}}
\caption{Allowed regions for $m_{H_2^+}$ {\it vs} $m_{H_1^+}$ from electroweak precision constraints on $\Delta T$ at the 1$\sigma$ [red (dark gray)] and 2$\sigma$ [green (light gray)] confidence levels. For $\beta = \pi/4$, $m_{A_0} =$~475~GeV and $m_{H_0} =$~70~GeV. The plus sign marks our benchmark point $(m_{H_2^+},m_{H_1^+}) = (850,90)$\;GeV (see body text for its full definition).}
\label{fig2:Cocktail}
\end{figure}
%%%%%%%%%%%%%%%%
%
In addition, the doubly charged scalar $\rho^{++}$ mediates lepton-flavor violation (LFV) at tree level in processes such as $\mu\rightarrow 3 e$ and
$\tau \rightarrow 3 e, 3\mu$, and at one-loop in processes like $\mu \rightarrow e \gamma$ and $\tau \rightarrow e \gamma,\mu \gamma$. This constrains the allowed values of $C_{ab}$ as a function of $m^2_{\rho}$, with the most stringent bounds being \cite{LFV,Hayasaka:2010np,Adam:2013mnn}
\be
\begin{array}{lrll}
\mu^-\rightarrow 3 e: 		&  | C_{e\mu}\,C_{ee} | 					& < & 2.3 \times 10^{-5}    \;(m_{\rho}/\mathrm{TeV})^2  \nonumber \\
\tau^-\rightarrow 3 e:    		&  | C_{e\tau}\,C_{ee} |  					& < & 9.0 \times 10^{-3}    \;(m_{\rho}/\mathrm{TeV})^2  \nonumber \\
\tau^-\rightarrow 3 \mu: 		&  | C_{\mu\tau}\,C_{\mu\mu} |				& < & 8.1 \times 10^{-3}    \;(m_{\rho}/\mathrm{TeV})^2  \nonumber \\
\tau^{-}\rightarrow \mu^{+} \,e^{-}\, e^{-}:    & | C_{\mu\tau}\,C_{ee} |			& < & 6.8 \times 10^{-3}   \;(m_{\rho}/\mathrm{TeV})^2  \nonumber \\
\tau^{-}\rightarrow \mu^{+} \,e^{-}\, \mu^{-}:	& | C_{\mu\tau}\,C_{e\mu} |		& < & 6.5 \times 10^{-3}   \;(m_{\rho}/\mathrm{TeV})^2  \nonumber \\
\tau^{-}\rightarrow e^{+}\, e^{-}\, \mu^{-}:   	& | C_{e\tau}\,C_{e\mu} |			& < & 5.2 \times 10^{-3}    \;(m_{\rho}/\mathrm{TeV})^2  \nonumber \\
\tau^{-}\rightarrow e^{+}\, \mu^{-}\, \mu^{-}:	& | C_{e\tau}\,C_{\mu\mu} | 		& < & 7.1 \times 10^{-3} \;(m_{\rho}/\mathrm{TeV})^2  \nonumber \\
\mu^+\rightarrow e^+ \gamma:		    	&\hspace{-4mm} \big|\sum_l C_{l\mu}\,C_{l e}^* \big|	& < & 3.2 \times 10^{-4} \;(m_{\rho}/\mathrm{TeV})^2.
\end{array}
\ee
LFV constraints favor $m_{\rho} \gtrsim 1$ TeV, which combined with Eq.\;(\ref{Constrainnumass}) leads to large values of $\kappa_2\gtrsim 1$ TeV and/or $\xi$,  close-to-maximal mixing $\beta \sim \pi/4$ and large mass splittings $\Delta m^2_{+},\Delta m^2_{0} \sim v^2$. For such large mass splittings, satisfying the EWPT constraints requires a mass spectrum $m_{H_{2,1}^+} \!\gtrsim\! m_{A_0} \!\gtrsim\! m_{H_{1,2}^+}$, resulting in a partial cancellation  of the $H_1^+$ and $H_2^+$ contributions in Eq.\;(\ref{deltaT}).

As a prototypical benchmark model that satisfies \mbox{EWPTs} (see Figure \ref{fig2:Cocktail}) and collider constraints, we take $m_{H_0} = 70$\;GeV, $m_{A_0} = 475$\;GeV, $m_{H_1^+} = 90$\;GeV, $m_{H_2^+} = 850$\;GeV and $m_{\rho} = 2$\;TeV, with $\kappa_2 = 3$\;TeV, $\xi=-2.5$ and $\beta = \pi/4$. Neutrino masses and mixings of the right size are then obtained for Yukawa couplings with absolute values: $C_{e\tau} \sim 0.32$, $C_{\mu\mu} \sim 0.066$, $C_{\mu\tau} \sim 3.6 \times 10^{-3}$ and $C_{\tau\tau} \sim 2.4 \times 10^{-4}$. Together with, {\it e.g.}, $C_{ee}\lesssim 10^{-2}$ and $C_{e\mu} \lesssim 10^{-3}$ these couplings fulfil at the same time all the LFV bounds. However, branching ratios for several LFV processes (like $\tau^{-}\rightarrow e^{+}\, \mu^{-}\, \mu^{-}$ and $\mu\rightarrow e \gamma$) are predicted close to the current experimental bounds, and may be probed in the near future.
% The contribution to the anomalous magnetic moments is tiny compared to current upper bounds, and would not ameliorates the current 3-4 sigma discrapency beteen the SM prediction and experiemntal findings.     

In this model, the short-distance contribution to neutrinoless double beta ($0\nu\beta\beta$) decays dominates over the one coming from light-neutrino exchange (since this one is proportional to $m_{ee}$ and thus suppressed by $x_e^2 \sim 10^{-9}$). If the value of $C_{ee}$ is not too small, this could open up the possibility to test this scenario at future $0\nu\beta\beta$ decay experiments. %\cite{Gustafsson:2014vpa}.

To conclude, we have put forward a minimal extension of the SM to include neutrino mass generation and dark matter in a unified framework, without introducing right-handed neutrinos. While giving an elegant explanation of the hierarchy $m_{\nu}/v$, the model predicts a small and nonzero value of $\theta_{13}$, together with a nontrivial relation between  $\theta_{13}$ and the octant of $\theta_{23}$, to be tested by future neutrino experiments. It also predicts LFV, WIMP dark matter with a mass of $\sim 50$-75~GeV and new scalar states to be searched for.

%\vspace{-4mm}
%\begin{acknowledgements}
%\vspace{-2mm}
%%%%%%%%%%%%%%%%%%%%%%%%%%%%%%%%%%%%%%%%%%%%%%%%%%%%%%%%%%%%
\bigskip
%\medskip
{\noindent \bf Acknowledgements.} We thank J.M.\ Fr\`ere, T.~Hambye, T.~Scarn\`a and M.~Tytgat for useful discussions, as well as  
S.~Borowka, G.~Heinrich and J.~Vermaseren for useful correspondence. 
M.A.R. is supported by Fondecyt Grant No. 11110472, Anillo ``Atlas Andino'' ACT1102 and DGIP Grant No. 11.12.39. J.M.N.\
is supported by the Science Technology and Facilities Council (STFC) under Grant No.\ ST/J000477/1.
M.G. is supported by the Belgian Science Policy (IAP VII/37), 
the IISN and the ARC project ``Beyond Einstein: fundamental 
aspects of gravitational interactions.''
%\end{acknowledgements}

%%%%%%%%%%%%%%%%%%%%%%%%%%%%%%%%%%%%%%%%%%%%%%%%%%%%%%%%%%%%
%%%%%%%%%%%%%%%%%%%%%%%%%%%%%%%%%%%%%%%%%%%%%%%%%%%%%%%%%%%%

\end{document}